\documentclass[12pt,preprint]{aastex}

\slugcomment{}

\shorttitle{X-ray emission from stellar bow shock}
\shortauthors{L\'opez-Santiago et al.}

\begin{document}


\title{AE Aurigae: first detection of non-thermal X-ray emission from \\
    a bow shock produced by a runaway star}


%
%
%
%
%
%
%

\author{J. L\'opez-Santiago\altaffilmark{1}, M. Miceli\altaffilmark{2,3}, 
M. V. del Valle\altaffilmark{4}, G. E. Romero\altaffilmark{4,5},            
R. Bonito\altaffilmark{2,3}, J. F. Albacete-Colombo\altaffilmark{6},
V. Pereira\altaffilmark{1}, E. de Castro\altaffilmark{1},
F. Damiani\altaffilmark{3}}


\altaffiltext{1}{Dpto. de Astrof\'isica y CC. de la Atm\'osfera, Universidad Complutense de Madrid, E-28040 Madrid, Spain}
\altaffiltext{2}{Dipartimento di Fisica, Universit\`a di Palermo, Piazza del Parlamento 1, 90134 Palermo, Italy}
\altaffiltext{3}{INAF-Osservatorio Astronomico di Palermo, Piazza del Parlamento 1, I-90134 Palermo, Italy}
\altaffiltext{4}{Instituto Argentino de Radioastronom\'ia (IAR), CCT La Plata (CONICET), C.C.5, 1894 
                      Villa Elisa, Buenos Aires, Argentina}
\altaffiltext{5}{Facultad de Ciencias Astron\'omicas y Geof\'isicas, Universidad Nacional de La Plata, 
                      Paseo del Bosque s/n, 1900 La Plata, Argentina}
\altaffiltext{6}{Universidad Nacional del COMAHUE, 
                      Monse\~nor Esandi y Ayacucho, 8500 Viedma, Rio Negro, Argentina}


\begin{abstract}
Runaway stars produce shocks when passing through interstellar medium
at supersonic velocities. Bow shocks have been detected in the mid-infrared for 
several high-mass runaway stars and in radio waves for one star. Theoretical 
models predict the production of high-energy photons by non-thermal radiative 
processes in a number sufficiently large to be detected in X-rays. {To date}, 
{no} stellar bow shock has been detected at such energies. {We present the 
first detection of X-ray emission from a bow shock produced by a runaway star}. The 
star is AE~Aur, which was likely expelled from its birth place by the encounter 
of two massive binary systems and now is passing through the dense nebula 
IC~405. The X-ray emission from the bow shock is detected at $30\arcsec$ 
to the northeast of the star, coinciding with an enhancement in the density of the 
nebula.
From the analysis of the observed X-ray spectrum of the source and our theoretical
emission model, we confirm that the X-ray emission is produced mainly by inverse 
Compton up-scattering of infrared photons from dust in the {shock front}. 

\end{abstract}





\section{Introduction}

{Runaway stars move through space with an extremely high velocity relative to the 
surrounding interstellar medium \citep{bla61}. 
%
A proposed} scenario for the ejection of stars at high velocities from their
birth place is close interaction between two binary systems \citep[e.g.][]{hoo00}. 
This process may produce ejection velocities close to the escape velocity 
at the surface of the most massive star in the system: up to $100-1000$ km\,s$^{-1}$
\citep{leo91,gva08}. 
{The encounter of the proper motions of two or more stars favors the} 
dynamical ejection scenario \citep{hoo00}. 
A typical example of a system of runaway stars escaping from a close interaction of two high-mass 
binaries is the group formed by AE~Aur, $\mu$~Col and $\iota$~Ori {\citep[e.g.][]{gua04}}.

The passage of a runaway star excites the interstellar medium forming a {shock front} 
\citep[e.g.][]{moh12} 
{which acquires the characteristic shape of a bow}. Several bow shocks
were identified in the past with the Infrared Astronomical Satellite 
\citep[IRAS;][]{van88, van95, nor97} and, more recently, with the \textit{Spitzer}
space telescope \citep{kob10} and other infrared missions \citep[e.g.][]{com07}.

Recently, \citet{ben10} detected non-thermal radio emission in the massive runaway star 
BD+43$^\mathrm{o}$~3654. This emission was associated with cooling of energetic
electrons through synchrotron radiation and the authors hypothesized that some bow shocks 
may also produce high-energy emission {that might be detectable.} 
This problem was treated theoretically in \citet{delvalle12}, 
{who predicted detectable X-ray and gamma-ray emission associated with particle 
acceleration at the bow shock}. No observational constraints were found to this 
phenomenon {to date}. 
{Very recently, \citet{ter12} reported the non-detection of X-ray emission from 
the bow shock produced by BD+43$^\mathrm{o}$~3654 in a long time exposure 
X-ray observation with \textit{Suzaku}.}
In this letter, we report the first detection of hard-energy emission from a bow shock. 


\section{The Runaway Star AE Aur (HIP~24575)}
\label{AEAur}

AE~Aur was ejected from its birth place in the Orion nebula cluster around
three million years ago, very probably by the encounter of two massive binary systems.
As a result, the stars AE~Aur and $\mu$~Col (both with spectral type 
O9.5) were expelled at high velocities, while the system $\iota$~Ori remained as 
a bounded binary system with high eccentricity \citep{hoo00}. 

AE~Aur {has recently} encountered IC~405, a dense molecular nebula with Hydrogen density 
$n_\mathrm{H} \sim 3$ cm$^{-3}$ \citep{per12}. As a result of this encounter, a bow shock 
was produced. The bow shock was detected in the mid-infrared band 
by \citet{van88} with IRAS and later by \citet{fra07} using \textit{Spitzer}.  
Some important parameters of the star for our study are the terminal velocity of the wind 
\citep[$v_\infty \approx 1500$~km\,s$^{-1}$;][]{hub11}, the mass-loss rate 
\citep[$\dot{M} \sim 10^{-7}$ M$_\odot$\,yr$^{-1}$;][]{ful06}, and the velocity of the star
\citep[$v_* \approx 150$~km\,s$^{-1}$;][]{per12}. {In Table~\ref{table}, we list a set 
of physical parameters of AE~Aur, including some assumptions on the characteristics
of relativistic particles adopted for our model.}


\begin{figure*}[!t]
\centering
\includegraphics[width=8.3cm]{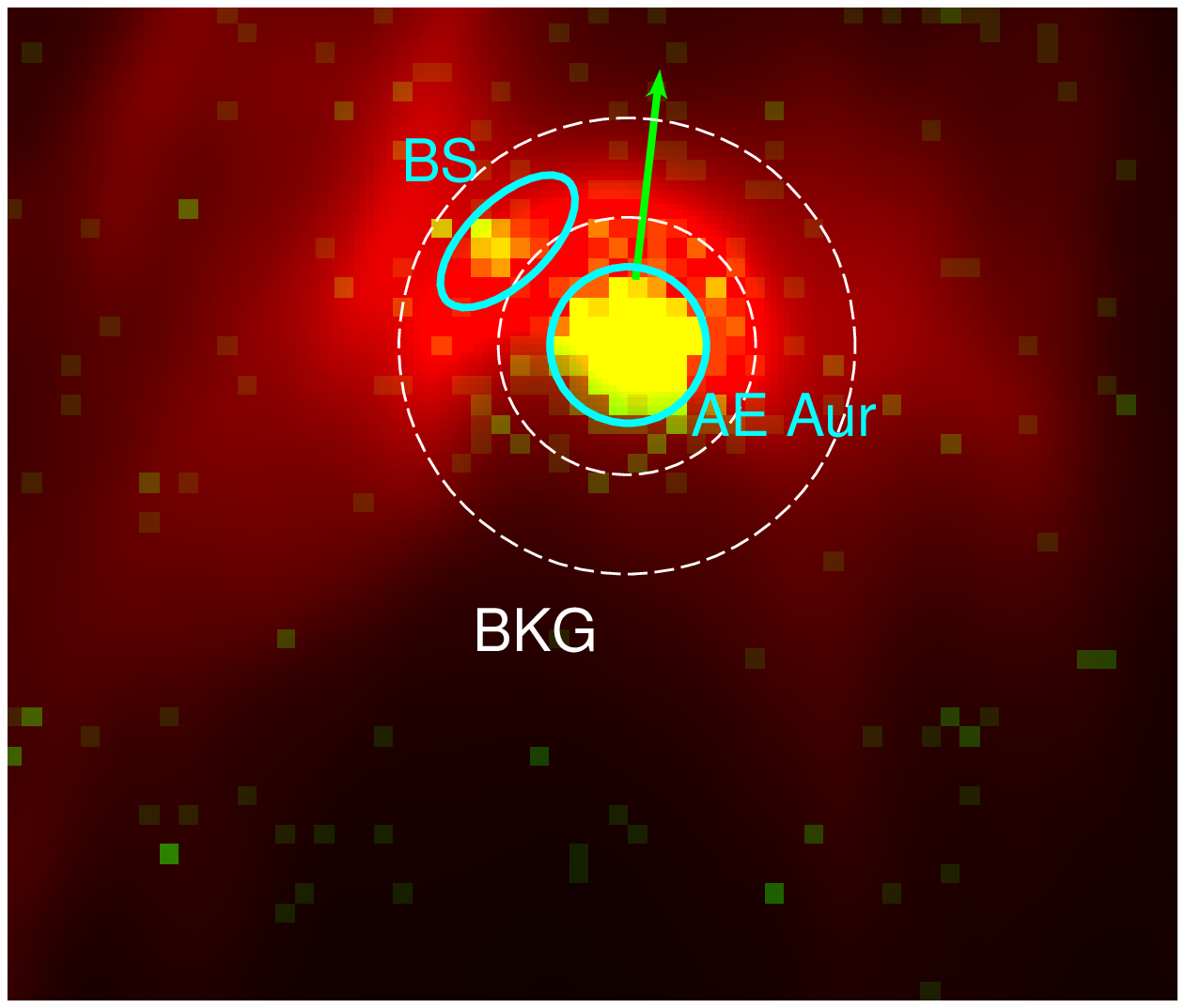}
\hspace{1cm}
\includegraphics[width=8.3cm]{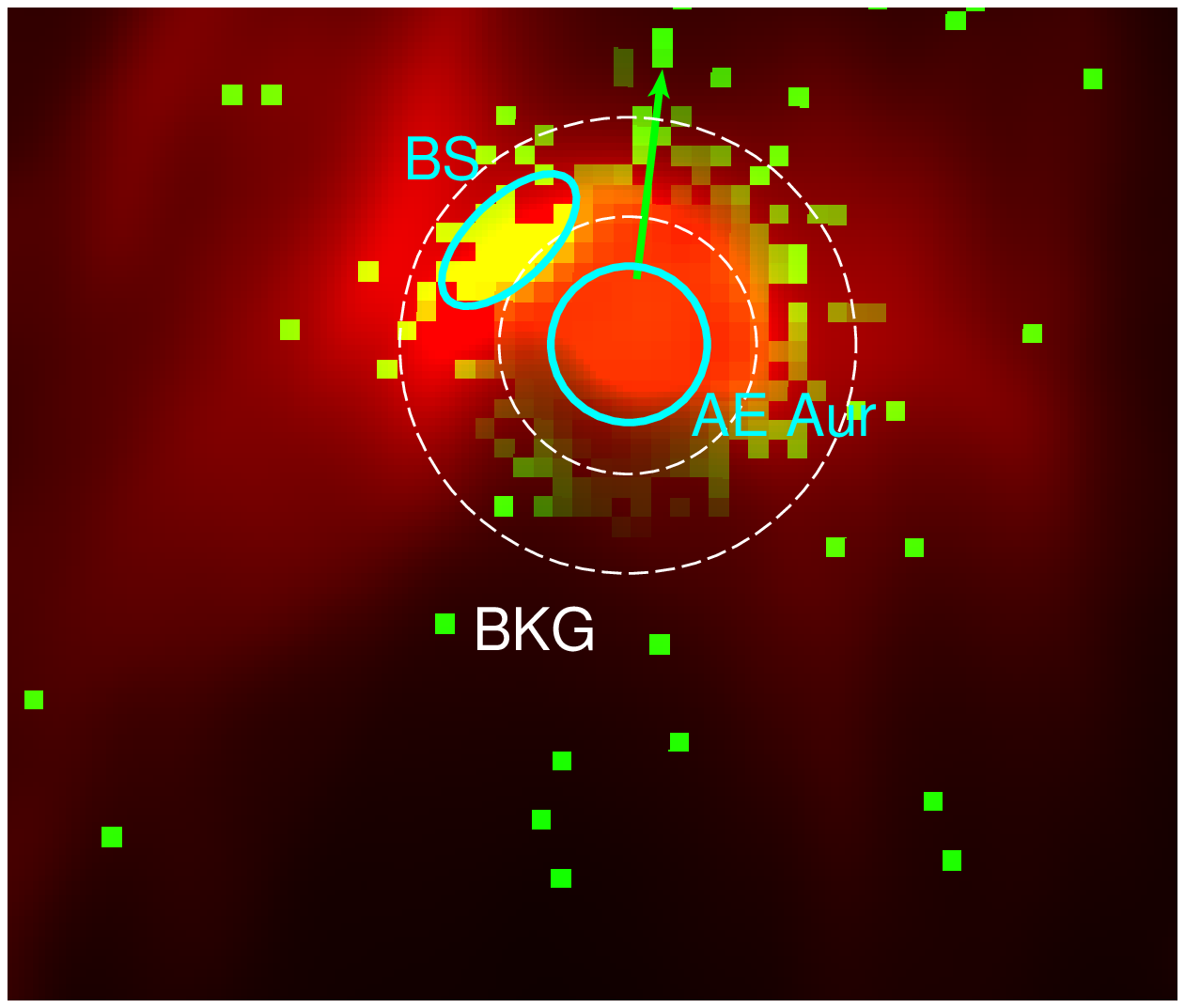}
\caption{\emph{Left panel}: WISE $12.1\mu$m image (red) and EPIC pn count image of AE Aur 
in the $1-8$ keV band (green). The bin size of the X-ray image is $4''$. The regions selected for 
the spectral analysis and the proper motion vector are superimposed. {BKG is the annular region 
from which the background spectrum was subtracted (see text).} North is up and East is to the left.  
\emph{Right panel:} same as left panel with pn median photon energy map in the $0.3-8$ keV in green. 
In the {median photon energy} map each pixel holds the median energy of the detected pn photons band. 
Only pixels with more than 4 counts were taken into account and the image is smoothed with $\sigma=8''$, 
according to the {formulae} in \citet{mdb08}.
\label{Ximage}}
\end{figure*}

\section{X-ray Observation and Data Analysis}
\label{Xray}

{For our work, we analyzed} the archive \emph{XMM-Newton} EPIC observation of AE Aur (ID 0206360101, PI F. Damiani),
performed on 2004 September 10 in Full Frame Mode with the Medium filter. The data were processed with the Science 
Analysis System (SAS V12.0). Light curves, images, and spectra, were created by selecting events with PATTERN$\le$12 
for the MOS cameras, PATTERN$\le$4 for the pn camera, and FLAG=0 for both. To reduce the contamination by soft 
proton flares, the original event files were screened by using the sigma-clipping algorithm (ESPFILT tasks). The screened 
MOS1$/$MOS2$/$pn exposure times are $52/54/38$ ks, respectively. Spectral analysis was performed in the $0.3-7$ keV 
energy band using XPEC V12.7 (the Ancillary Response Files were generated with the SAS $ARFGEN$ task). 
MOS and pn spectra were rebinned to achieve a signal-to-noise ratio per bin $>5\sigma$. All the reported errors are 
at the 90\% confidence level.

Figure \ref{Ximage} shows the WISE $12.1\mu$m image (in red, see also \citealt{per12}) of the bow shock of AE Aur together 
with the corresponding EPIC pn count image in the $1-8$ keV band (left panel, in green) and the pn median photon energy map, 
i.e., an image where each pixel holds the median energy of the detected pn photons in the $0.3-8$ keV band 
(right panel, in green; for details on the procedure adopted to produce the map, see \citealt{mdb08}). A bright X-ray source, 
$\sim 30''$ northwest of the star is visible in the left panel of Fig. \ref{Ximage} (BS region) and appears embedded 
in the infrared bow-shock. We verified that BS has {no point-like} counterparts in the optical and infrared bands. The 
{median photon energy} map
clearly shows that the X-ray emission from the BS source is significantly harder than that of AE Aur\footnote{We verified that 
the MOS cameras provide consistent results. Here we present pn images because of the better statistics}. Namely, the 
average value of {median photon energy} is $\sim850$ eV in region BS and $\sim750$ eV in region AE Aur. Moreover, 
{this map reveals} a bow shaped structure (not directly visible in the count image) characterized by hard X-ray emission and 
somehow reminiscent of the infrared bow shock.

\begin{figure}[!t]
\centering
\includegraphics[width=9.5cm]{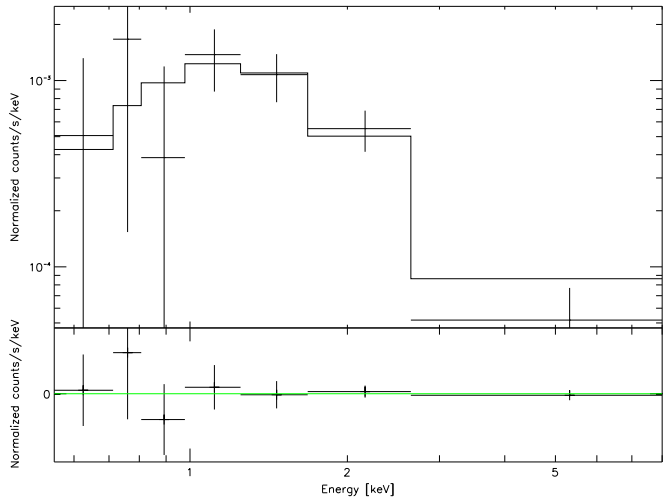}
\caption{Background-subtracted EPIC pn spectrum of region BS 
of Fig. \ref{Ximage} with the theoretical model described in 
Sect.\ref{nonthermal} and the corresponding residuals 
{(Observed - Model)}.
\label{Xray}}
\end{figure}

We first analyzed the stellar spectrum by extracting the pn and MOS spectra from region ``AE Aur" of Fig \ref{Ximage}. We
subtracted to the source spectrum a background spectrum extracted from a nearby region {(out of the field of view 
of Fig \ref{Ximage}).} 
The fittings were performed simultaneously on both MOS spectra and on the pn spectrum by adopting an absorbed APEC 
model (optically thin coronal plasma in collisional ionization equilibrium, \citealt{sbl01}) based on the 2.0 release of the 
AtomDB database. The interstellar absorption was described by the PHABS model in XSPEC. The stellar spectrum can 
be fitted (reduced $\chi^2=1.32$, with 315 d.~o.~f.) by an isothermal plasma with temperature $kT=0.225^{+0.007}_{-0.004}$ 
keV, emission measure $EM=2.8^{+0.5}_{-0.4}\times10^{55}$ cm$^{-3}$ (assuming a distance $D=550$ pc, in agreement 
with \citealt{per12}), and oxygen abundance\footnote{We verified that the plama metallicity is $\sim0.35$ and we fixed the
abundances of all other elements to this value.} $O/O_\sun =0.26\pm0.02$. The best-fit value of the column density is
$n_H=4.2\pm0.2\times10^{21}$ cm$^{-2}$ and the unabsorbed X-ray luminosity in the $0.3-10$ keV band is 
$L_\mathrm{X} = 1.9 \times 10^{32}$ erg\,s$^{-1}$.

We then analyzed the X-ray spectrum of the BS source (the extraction region is shown in Fig.~\ref{Ximage}). The contamination
from AE~Aur photons scattered by the telescope PSF in the BS region was removed by extracting the background spectrum 
from the region BKG of Fig. \ref{Ximage}. 
{This region consists of an annulus centered at the star and with inner and outer radii that were chosen to contain pixels 
at the same distance from the star than the bow shock\footnote{Pixels inside the BS region were extracted from the background 
extraction region (BKG).}.}
Figure \ref{Xray} shows the pn spectrum of region BS. {Only 180 pn counts (943 ph\,arcmin$^{-2}$)} were detected in 
this region, while the spectra from the MOS cameras do not provide useful constraints and are not presented here. 
{The BS spectrum can be well fitted by an absorbed power law model with photon index $\Gamma=2.6^{+0.6}_{-0.5}$ 
and normalization $N=6\pm2\times10^{-6}$ photons keV$^{-1}$ cm$^{-2}$ s$^{-1}$ at 1 keV.} Since the absorbing 
column is consistent with that found for AE Aur, we fixed it to $n_\mathrm{H} = 4.2 \pm 0.2 \times 10^{21}$ cm$^{-2}$ 
in the fittings, to obtain tighter constraints on the other model parameters.
{We used Cash instead of $\chi^2$ Statistics for the fitting, that is recommended when the number of counts in a 
channel is small. We used the task \textit{GOODNESS} with 1000 Montecarlo simulations to determine the goodness of 
our fit. The value obtained for our fit is 2.3\% of realizations with less best fit statistic than that obtained with the fitted model
(i.e. 97.7\% confidence interval).}

The BS spectrum can also be well fitted by a very hot thermal component 
with $kT=2.4^{+2}_{-0.8}$
keV and $EM=4\pm1\times10^{52}$ cm$^-3$ (assuming that BS is at $550$ pc). Such extremely high temperature cannot be
associated with the stellar wind, as shown below. If we assume that the hot plasma fills the whole region, we derive from the 
$EM$ a lower limit for the particle density, $n\sim3$ cm$^{-3}$, that is much higher than that expected for a stellar 
wind\footnote{If the plasma does not fill the entire region we get even higher density values.}. Temperatures higher than 1.6 keV
can be indicative of a cataclysmic variable star or of a very energetic coronal flare, though we verified that the BS light curve 
is {consistent with a constant source}. Given the low count-rate of the source, statistical fluctuations are quite large and
deeper exposure are necessary to obtain a proper study of the light curve. In conclusion, present data do not allow us to
discriminate between a thermal and a non-thermal scenario for the BS X-ray emission and deeper observations are necessary.
Nevertheless, the lack of infrared and optical counterparts and the extremely high temperature derived from the thermal model
make the association of BS with a foreground$/$background stellar object less likely. On the contrary, the spatial correlation
between BS and the infrared bow shock, the bow-shaped morphology of the hard X-ray emission around AE Aur, and the 
spectral properties of BS (see next section) concur in suggesting that BS is indeed the result of non-thermal emission 
originating at the bow shock of the runaway star AE Aur.

\section{Non-Thermal Radiative Model}
\label{nonthermal}

In order to explain the observed X-ray emission {from the bow shock of AE~Aur}, we have applied 
the non-thermal radiative model {of $\zeta$ Oph,} developed by \citet{delvalle12} {with the parameters given in 
Table~\ref{table}  (see also $\S$~\ref{AEAur})}. {By adopting stellar parameters of an O9.5V star and a 
ISM medium $N_\mathrm{H} =  3$\,cm$^{-3}$ \citep{per12}, we obtain that}
the runaway star produces a bow shock in the interstellar medium with a stagnation point 
located at approximately $17.1 \times 10^{3}$ AU. {This estimation agree very well with the value 
measured from the X-ray observation ($\sim 17 \times 10^{3}$ AU).} 
The reverse shock in the stellar wind is adiabatic and provides 
an efficient site for particle acceleration through a Fermi~I {type acceleration} process. Since the gas is 
compressible, we assume that it is in sub-equipartition, getting an estimate of $\sim 10^{-4}$ G 
{for the magnetic field in this medium} \citep[see][for details]{delvalle12}. 

\begin{table}
\centering
\caption[]{Parameters for the non-thermal radiative model for AE~Aur}
\begin{tabular}{lll}
\hline\noalign{\smallskip}
\multicolumn{2}{l}{Parameter} & value\\
\hline\noalign{\smallskip}
$R_{\rm 0}$ & Standoff radius & 1.7 $\times 10^{4}$ AU \\
$\dot{M_{\rm w}}$ & Wind mass loss rate& 10$^{-7}$ M$_{\odot}$ yr$^{-1}$\\
$a$ &Hadron-to-lepton energy ratio & 1 \\
$q_{\rm rel}$ & Content of relativistic particles & 0.007   \\
$\alpha$ &Particle injection index & 2.6\\
$V_{\rm w}$ & Wind velocity & 1.5$\times10^{8}$ cm s$^{-1}$   \\
$L$ & Available power & 4$\times10^{33}$ erg s$^{-1}$\\
$B$ &Magnetic field & 1.1$\times 10^{-4}$ G  \\
$V_{\star}$ & Star velocity &  150 km s$^{-1}$ \\
\hline\\
\end{tabular}	
  \label{table}
\end{table}

The available power in the stellar wind is $\sim 4\times10^{33}$ erg\,s$^{-1}$. {Assuming that only}
a small fraction of $\sim 1$ \% is transformed in relativistic particles (we assume equal power in 
leptons and hadrons), {we have been able} to reproduce the observed X-ray spectrum with a power-law electron 
injection of index $-2.6$, through inverse Comptom up-scattering of IR photons coming {from} dust that is heated 
by the forward shock and the thermal radiation. 
{This model produces a good fit to the observed spectrum of the bow shock (see Fig.~\ref{Xray}), 
with reduced $\chi^{2} = 0.43$ with 5 d.o.f. (see $\S$~\ref{Xray}). 
{Models with higher ratio of proton to electron power are possible. Since the cooling time of the protons is 
much longer, they are removed from the radiation zone without having time to make a significant contribution to the 
spectral energy distribution. So, in order to keep the goodness of the fit, a larger total power is required for 
proton-dominated models. In a case similar to that of the Galactic cosmic rays, where the mentioned ratio is 100 
instead of 1, we should consider that about 0.3 of the available power goes to relativistic particles. This requires very 
efficient acceleration. The case with $a=1$ is energetically more conservative, but we cannot rule out models with $a$ 
up to 100. In all cases, the shape of the SED is basically the same and determined by the electron population.
}

In Figure~\ref{model}, we present the broadband spectral energy distribution {implied by our model. Inverse Compton 
losses dominate, producing} a steepening of the non-thermal spectrum that results in the soft power-law X-ray emission 
observed. The dominance of the radiative losses requires a low level of convection of the 
relativistic particles that can result from the development of Kelvin-Helmholtz instabilities in the bow shock 
\citep[see conditions for these instabilities in bow shocks in][]{ara10}.

\begin{figure}[!t]
\centering
\includegraphics[width=5.75cm,angle=270]{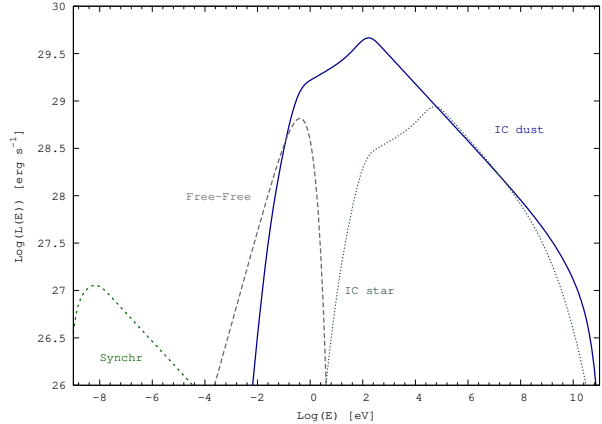}
\caption{Spectral Energy Distribution of non-thermal radiative processes for the AE~Aur bow shock. 
The model was computed following the work by \citet{delvalle12}. The continuous line 
represents the inverse Compton process produced by the dust (See $\S$~\ref{nonthermal} for 
details).
\label{model}}
\end{figure}

The SED predicted for the runaway star has little and steep synchrotron radio emission and a peak of around 
$\sim 10^{29.5}$ erg s$^{-1}$ at soft X-rays. Contrary to the case of $\zeta$ Oph, located closer and with different 
physical parameters, the expected emission at gamma-rays is negligible. Hence, the best test for the proposed model 
is the radio detection through deep interferometric observations at long {centimetric} wavelengths.

\section{Conclusion}

{We detected a hard X-ray source spatially correlated with the infrared bow shock of the runaway star AE Aur. 
The source is consistent with being a point-like steady source with non-thermal emission, though deeper observations 
are necessary to ascertain its morphology, its variability (if any),  and its {photon index}. 
Nevertheless, the analysis of the X-ray source at $\sim 30\arcsec$}
to the northeast of the runaway star AE Aur
showed that it is very likely the X-ray counterpart of the bow shock produced by the passage 
of the star through the dense molecular nebula IC~405 and that was detected in the infrared
with {IRAS, \textit{Spitzer} and WISE.} We developed a radiative non-thermal model 
specifically tuned to describe the AE Aur bow shock 
properties and obtained a very good agreement between model and observations.
According to our model, the X-ray emission would be produced by inverse Compton 
process of accelerated particles in the nebula dust. High-energy emission from bow shocks 
produced by runaway stars has been predicted by theoretical models. This is the first time that 
X-ray emission is detected in one of such bow shocks. 
Future X-ray and radio observations of 
this source will permit to further constrain its properties.

\acknowledgments

This work was supported by the Spanish Government  (AYA2011-30147-C03-02 and 
AYA2011-29754-C03-03). J.L.-S. thanks project AstroMadrid (S2009/ESP- 1496) for 
partial support. G.E.R. and M.V.dV. were supported by PIP 0078 (CONICET) 
and PICT 2007-00848, Pr\'estamo BID (ANPCyT). G.E.R. received additional support 
from the Spanish Ministerio de Inovaci\'on y Tecnolog\'ia (AYA 2010-21782-C03-01).
{We would like to thank the referee for useful comments and suggestions.}



{\it Facilities:} \facility{XMM-Newton (EPIC)}.

\clearpage

\end{document}